# Extreme electrodynamics in time-varying media

M. Scalora[1], M. A. Vincenti[2], D. de Ceglia[2], N. Akozbek[3], M. Ferrera[4], C. Rizza[5], A. Alù[6,7], N. Litchinitser[8], C. Cojocaru[9] and J. Trull[9]

[1]*FCDD-AMT-MGR, DEVCOM AvMC, C. M. Bowden Research Center, Redstone Arsenal, Alabama, 35898-5000, USA*
[2]*Department of Information Engineering – University of Brescia, 25123 Brescia, Italy*
[3]*US Army Space & Missile Defense Command, Tech Center, Redstone Arsenal, AL 35898 USA*
[4]*Institute of Photonics and Quantum Sciences Heriot-Watt University, SUPA Edinburgh, EH14 4AS United Kingdom*
[5]*Department of Physical and Chemical Sciences, University of L'Aquila, I-67100 L'Aquila, Italy*
[6]*Photonics Initiative, Advanced Science Research Center, City University of New York, New York, NY 10031, USA*
[7]*Physics Program, Graduate Center, City University of New York, New York, NY 10016, USA*
[8]*Department of Electrical and Computer Engineering, Duke University, Durham, NC, 27708, USA*
[9]*Department of Physics, Universitat Politècnica de Catalunya, 08222 Terrassa (Barcelona), Spain*

**Abstract**

Abrupt time variations of the properties of optical materials have been at the center of intense research efforts in recent years, with the prospect of enabling extreme wave transformations and of leveraging time as a degree of freedom for wave control. While the most viable approach to yield ultrafast variations of the optical material response is through optical pumping of nonlinear media, the complex dynamics in these systems are not yet fully understood. Here, as a relevant case study, we rigorously investigate the pump-probe dynamics in a 310nm-thick transparent conductive oxide etalon, using a weak 40 femtosecond probe and a pump that displays peak power densities in the TW/cm$^2$ range with a duration of a few femtoseconds. We examine the pump-probe interaction using a hydrodynamic-Maxwell approach that accounts for diffraction, self-focusing and -defocusing, self- and cross-phase modulation, probe gain, and linear and nonlinear material dispersion expanded in the perturbative regime up to 9$^{th}$ order for both pump and probe. By allowing the intricacies of the pump-probe interaction to proceed in time, we can also define an effective spatio-temporal permittivity for a more direct evaluation of the material ultra-broadband optical behavior. The reported results challenge the conventional modeling of this kind of problem, which has so far overlooked pump dynamics, simplistically assigning a local time-dependent refractive index to the probe that may be designed to fit the experimental data, but has no physical connection to the complex pump-probe interaction. Our approach unveils new dynamics, pointing towards the possibility to achieve extreme pulse compression into the attosecond range and nonlinear diffraction over deeply subwavelength propagation distances, thus opening a possible new path towards novel and cost-effective tools for integrated photonics and attosecond science.

**Introduction**

Spatio-temporal or 4-dimensional metamaterials have been in the spotlight in recent years, holding the promise to sustain novel forms of light-matter interactions and enhanced functionalities based on harnessing phenomena arising at the nanoscale in the ultrafast, single cycle regime [1-9]. The growing interest in this field, spanning from the optical to the RF and acoustic communities, is due to recent technological advances in introducing and controlling the material responses at fast timescales in planarized, ultrathin devices. The possibility to control electromagnetic waves using such flat devices [10], with deeply subwavelength thicknesses and tailored spatial properties, is attractive for several reasons: (1) realization of cost-effective devices with reduced sizes; (2) compatibility with integrated circuits; (3) phase and amplitude control; (4) ease of tunability. Frequency-selective surfaces [11] and metal-mesh filters [12] are popular examples of such artificial surfaces that predate the more recent excitement around metasurfaces [13-15]. Adding a temporal element [16-18] to this planarized platform for wave control can have game-changing consequences in terms of broader applications, functionalities and performance in the same way that photonic band gap structures have revolutionized photonic devices [19]. For example, time-variations can enable parametric phenomena that overcome passivity and bandwidth constraints of passive, linear, time-invariant devices [20]. While current theoretical approaches and fabrication technologies to model and control the electromagnetic response of materials may be adequate for RF/microwave applications, optical regimes offer new challenges, both from the theoretical and the experimental standpoint.

Processing light in nanoscale components entails a detailed description of the field-medium interaction beyond the one given by classical macroscopic electrodynamics. The dynamics of atomic, molecular or condensed matter media must be considered from a Maxwellian point of view, with wave propagation in macroscopic media using either a hydrodynamic-Maxwell approach, a Bloch-Maxwell formulation (non-perturbative components), or a combination of both, depending on material composition, doping and size. For high intensities, single-cycle pulses and nanoscale size objects, which emerge in the realization of space-time metamaterials at optical frequencies, common approximations that are adopted in the macroscopic theory, like the slowly varying or rotating wave approximations, may no longer be valid as new linear and nonlinear optical phenomena are triggered. Some important examples are nonlocal effects arising from electron gas pressure and viscosity that make the dielectric response a simultaneous function of frequency and wave vector [21-26]; electron cloud screening [27]; quantum tunneling [28, 29]; negative refraction and related magnetic phenomena [30]; filamentation and plasma formation [31]; bound electron contributions to the dielectric response [32]; orbital angular momentum [33]; near dipole-dipole interactions [34, 35]; local field effects in dense media [36]; high harmonic generation [37, 38]; a modern reformulation of damping that supplants the Abraham-Lorentz theory [39-42]; resonant, multi-level component that may be present as dopants [43] that can enhance either electric or magnetic field responses.

Progress in nanotechnology has produced devices with near-atomic size. An optical material is generally composed of Drude and Lorentz oscillators both exhibiting nonlinearities that must be expanded up to an unspecified order to accurately describe the microscopic perturbative response [44]. These charges may interact with incident fields only a few cycles in duration and have peak power densities of order 10 TW/cm$^2$ or more, in either single incident pump pulse configuration for harmonic generation [37, 38], or pump-probe arrangements [3] for beam control and optical processing. The electronic response may be overwhelmed by thermal effects, which may shift the Fermi surface or change the effective masses of free and bound electrons. These changes affect harmonic generation, which is sensitive to the symmetry properties of the material [45]. Free electrons in turn can shield the bulk ones and affect light-matter interactions. The combination of screening, shifting Fermi surfaces, boundary conditions, including geometrical resonances, possible epsilon-near-zero (ENZ) response, tunneling if material components are placed at distances of 1 nm or less, can likely reveal novel phenomena. Hence the need for adequate, accurate models.

With reduced sizes, the level of required accuracy increases demands on design tools and algorithms that are used to predict device performance. The complexity required to describe a pump-probe interaction with matter that proceeds in extreme scenarios (i.e., intensities exceeding 1 TW/cm$^2$ and/or few-cycle pulsed excitation) calls for a rigorous approach that includes writing equations of motion for the polarizations at the two frequencies coupled to Maxwell's equations, with the strong pump field driving the much weaker probe field. By contrast, the standard approach that has been used so far in modeling these optically pumped materials experiencing ultrafast changes in optical response [1-9, 16, 17, 46-49] has been to disregard pump dynamics entirely, and to ignore all external factors that may lead to internal material changes, like red- or blue-shift of the plasma frequency, increased free electron charge density because of interband transitions, and dynamic shifts of the ENZ crossing point for free electron and resonance frequency for bound electrons. Instead, an artificial and arbitrary temporal modulation of the dielectric function has been commonly adopted to drive the probe in a way that only marginally mimics the effects of the pump, while yielding a probe response that may approximately conform with observations. For example, in the case of transverse plane wave excitations, the standard approach is to assume that $\epsilon_r(t,z) = \epsilon_{TCO} + \Delta\epsilon(t)$, where $\Delta\epsilon(t)$ is determined by the pump interaction with the material. This definition often assumes an instantaneous and dispersionless response [46], requires a low-density medium [46-49] and is used to justify a transformation of the displacement vector into the form $D(t,z) = \varepsilon(t,z)E(t,z)$, which finds its way into Maxwell's equations for the probe.

In addition to possible issues associated with causality [46-49] due to the arbitrary nature of $\Delta\varepsilon(t)$, and challenges related to the ability to achieve required power levels for adequate temporal modulation [49], the use of a basic function like $\varepsilon_r(t,z)$ under extreme conditions, where dispersion and absorption are not negligible, inevitably leads to neglecting pump dynamics. This implies disregard of effects like self-phase modulation, self-steepening, self-focusing and de-focusing, diffraction, intra-pulse Raman scattering, possible probe gain, etc. The temporal modulation of the dielectric response $\Delta\varepsilon(t)$ contributes to an erroneous description of the pump, which leads to an equally inaccurate description of the probe. As an example, in the Supplementary Information we address issues related to the speed of light of both pump and probe under extreme conditions and show that the probe's energy velocity is a function of peak power density.

In view of the general trend toward extreme nonlinear effects triggered by ultra-intense and ultra-short optical pulses in highly dispersive systems (e.g., transparent conductive oxides, TCOs, operating close to their ENZ frequency), we focus this effort on studying the extreme nonlinear optical response in these systems in a way that may enable accurate predictions of spatio-temporal light-matter interactions, thus facilitating the modeling and interpretation of experimental observations at interfaces and nanostructures composed of metals, semiconductors and TCOs. In this regard, a good testbed for our model is provided by the experimental results reported in [3], which to date probably represent the most extreme case of optical nonlinearities triggered by near-single cycle pulsed excitations in TCO thin films. In [3], a TM-polarized pump pulse 6 fs in duration having peak power density of 2 TW/cm$^2$ and tuned to 800nm is normally incident on a 310 nm thick Indium Tin Oxide (ITO) layer. This pulse is then used to modulate a much weaker

TM-polarized probe pulse approximately 40 fs in duration incident at near-normal angle (we assume 2°) and tuned to 1200 nm, which is near the ENZ crossing point.

**The Hydrodynamic-Maxwell Model**

The macroscopic electrodynamics of TCOs like ITO, AZO, and doped cadmium oxide (Indium doped CdO, In:CdO, or Dysprosium doped CdO, Dy:CdO) are theoretically described starting with the ellipsometric retrieval of the *local* dielectric response of the sample. This frequency-dependent, *linear, complex* function is then fitted using a Drude-Lorentz model with a resonance typically found at ultraviolet (UV) wavelengths, extending into the visible range, and by combining it with the Drude contribution that extends into the infrared (IR) range to yield the expression:

$$\varepsilon_{TCO}(\omega) = 1 - \frac{\omega_{p,f}^2}{(\omega^2 + i\gamma_f \omega)} - \frac{\omega_{p,b}^2}{(\omega^2 - \omega_{0,b}^2 + i\gamma_b \omega)} \qquad (1)$$

The presence of additional bound electron species may be accounted for by simply adding Lorentzian functions to Eq.(1). Here, $\omega_{p,f}$, $\omega_{p,b}$ are the constant free and bound electron plasma frequencies; $\omega_{0,b}$ is the resonance frequency; and $\gamma_f$, $\gamma_b$ are the free and bound electron damping coefficients. Fitting the retrieved local dielectric constant with available data allows us to determine damping coefficients, effective masses, and densities, which are inserted into material equations of motion to replace Eq.(1) and study the system's dynamics. The classic hydrodynamic model considers only the free electron response. In the following, we present key aspects of a hydrodynamic-Maxwell model that naturally incorporates causality, linear and nonlinear dispersions, free and bound electron dynamics in a self-consistent manner and makes no *a priori* assumptions about the spatio-temporal forms of the plasma frequency, effective mass, number density, or the time dependence of the index of refraction. In addition, any shifts of the bound electron resonance frequency and the ENZ crossing point occur because of the interaction between fields, and not because of any artificial modulation that may appear to violate electromagnetic energy and momentum conservation. The model has been used previously to describe second- and third-harmonic generation in an ITO nanolayer [50], and high-harmonic generation in both AZO [37] and silicon [38]. With appropriate modifications for pump-probe dynamics, which amount to retention of terms that contribute only to the fundamental pump and probe wavelengths, the equations of motion are given by:

$$\ddot{\mathbf{P}}_f + \tilde{\gamma}_f \dot{\mathbf{P}}_f = \frac{n_f(\mathbf{r},t)e^2\lambda_r^2}{m_f^*(T_e)c^2}\mathbf{E} + \frac{3E_F}{5m_f^*(T_e)c^2}\left[\nabla(\nabla\bullet\mathbf{P}_f) + \frac{1}{2}\nabla^2\mathbf{P}_f\right] \quad (2)$$

$$\ddot{\mathbf{P}}_b + \tilde{\gamma}_{0b}\dot{\mathbf{P}}_b + \tilde{\omega}_{0b}^2\mathbf{P}_b + \mathbf{P}_{b,NL} = \frac{n_{0b}e^2\lambda_r^2}{m_b^*c^2}\mathbf{E} \quad (3)$$

Equations (2) and (3) describe the dynamics of free and bound electrons, respectively, where $m_f^*$ and $T_e$ are the free electron mass and its temperature. The nonlinear polarization of the bulk crystal is expressed as:

$$\mathbf{P}_{b,NL} = -\beta_1(\mathbf{P}_b\bullet\mathbf{P}_b)\mathbf{P}_b + \beta_2(\mathbf{P}_b\bullet\mathbf{P}_b)^2\mathbf{P}_b - \beta_3(\mathbf{P}_b\bullet\mathbf{P}_b)^3\mathbf{P}_b + \beta_4(\mathbf{P}_b\bullet\mathbf{P}_b)^4\mathbf{P}_b + \dots \quad (4)$$

Equation (4) reflects the fact that media like TCOs and noble metals are centrosymmetric, so that even powers may be neglected in the expansion. The spatial and temporal derivatives are expressed in the scaled coordinate system as: $\tau = ct/\lambda_r$ is time, $\zeta = x/\lambda_r$ and $\varsigma = y/\lambda_r$ are transverse coordinates, and $\xi = z/\lambda_r$ is the longitudinal coordinate; $\lambda_r = 1\mu m$ is the reference wavelength. $m_b^*$ is the bound electron mass; $n_f(\mathbf{r},t)$ is the free electron density; $n_{0,b}$ is the constant bound electron density; $E_F = \hbar^2(3\pi^2 n_{0,f})^{2/3}/2m_0^*$ is the Fermi energy; $\tilde{\gamma}_{f,b} = \gamma_{f,b}\lambda_r/c$ and $\tilde{\omega}_{0,b}^2 = \omega_{0,b}^2\lambda_r^2/c^2$. In the Supplementary Information we discuss circumstances where the free electron density in the conduction band can increase via interband transitions, and where both free electron damping and pressure become functions of temperature and consequently functions of the field intensity. The coefficients in the nonlinear background polarization Eq.(4) are $\beta_j = \omega_{0,bj}^2\lambda_r^2/c^2(d^2 n_{0,b}^2 e^2)^j$, where $d$ is either the lattice constant or the size of the orbital participating in the process. They can be derived from a classical oscillator model, may be dispersive [38], and are to be interpreted as tensors that contain information about crystal symmetry. Without loss of generality, here we assume that third- and higher-order nonlinear coefficients reflect the symmetry of an isotropic medium, though they can be modified to reflect other crystal symmetries. The combination of these equations preserves linear and nonlinear dispersion effects, while the polarization in our model is expanded up to the 9[th] order due to the high peak power densities of the fields. Incident pulses may have either ultrawide bandwidth components [3] that may undergo strong self-phase modulation and spectral broadening, or be tuned in the Lorentzian range [51] to trigger interband transitions. It is important to stress that, from a numerical standpoint, an intrinsic and consistent instability is observed for large peak power densities if the expansion is not developed to a sufficiently large order. The free electron component in a TCO is characterized by an effective electron mass $m_f^*$ which typically is temperature dependent. The second term on the right-hand side of Eq.(2) represents the usual linear nonlocal contributions, i.e., pressure and viscosity, which become important for thin layers.

Hot carriers are typically handled using the two-temperature model [52, 53], which couples lattice and electron gas temperatures to determine the instantaneous plasma frequency, carrier density, damping coefficient, and ultimately can be used to express the temperature dependence of the effective mass. However, the pulse durations that we are exploring are unusually short to contemplate the kinds of temperature effects outlined above [54]. Instead, free electrons are energized by the absorption of laser energy, adjusting instantaneously to the shape of the laser pulse. Electron temperature rises sharply, while the lattice remains cold. As a result of absorption, the electrons occupy different states close to the Fermi level, imparting a different effective mass to the electrons. Therefore, focusing on the near-instantaneous character of the interaction, we can then express the effective mass as a linear function of absorption:

$$m_f^* \approx m_0^* + \Lambda \iint \mathbf{J} \cdot \mathbf{E} d\mathbf{r}^3 dt \quad , \quad (5)$$

where $\Lambda$ is a constant of proportionality; $m_0^*$ is the electron's rest mass; $\Lambda \iint \mathbf{J} \cdot \mathbf{E} d\mathbf{r}^3 dt$ represents absorption; $\mathbf{J} = \dot{\mathbf{P}}_f$ is the current density, which is readily available as the solution of Eq.(2). Keeping it inside the integral of Eq.(5) would properly account for retardation effects and slower components of the nonlinearity. However, for simplicity, in the present effort we may assume that $\mathbf{J} = \sigma_0 \mathbf{E}$, where $\sigma_0$ is to be determined, an assumption that is equivalent to retaining only the ultrafast, instantaneous response. Using the expression for the effective mass, the leading terms in Eq.(2) become

$$\frac{n_{0,f} e^2 \lambda_r^2}{m_f^* c^2} \mathbf{E} = \frac{n_{0,f} e^2 \lambda_r^2}{m_0^* c^2} \left( 1 - \frac{\Lambda \sigma_0}{m_0^*} \iint \mathbf{E} \cdot \mathbf{E} d\mathbf{r}^3 dt + \left( \frac{\Lambda \sigma_0}{m_0^*} \iint \mathbf{E} \cdot \mathbf{E} d\mathbf{r}^3 dt \right)^2 - \ldots \right) \mathbf{E} . \quad (6)$$

We can make one further assumption to simplify the integrals in Eq.(6) by introducing the parameter $\tilde{\Lambda} = \Lambda \sigma_0 V \tau_0 / m_0^*$ where $V$ is the interaction volume and $\tau_0$ is the temporal duration of the pulse. We recognize that $\Lambda$, $\sigma_0$, and $m_0^*$ depend on wavelength, band curvature, and fluence, so that $\tilde{\Lambda}$ is to be understood as a parameter that combines information about fluence, conductivity, dispersion, and determines the spatio-temporal dynamics of the redshift in the plasma frequency. Eq.(6) then takes the simplified form:

$$\frac{n_{0,f}e^2\lambda_r^2}{m_f^*(T_e)c^2}\mathbf{E} = \frac{n_{0,f}e^2\lambda_r^2}{m_0^*c^2}\mathbf{E} - \tilde{\Lambda}(\mathbf{E}\cdot\mathbf{E})\mathbf{E} + \tilde{\Lambda}^2(\mathbf{E}\cdot\mathbf{E})^2\mathbf{E} - \tilde{\Lambda}^3(\mathbf{E}\cdot\mathbf{E})^3\mathbf{E} + \tilde{\Lambda}^4(\mathbf{E}\cdot\mathbf{E})^4\mathbf{E} \quad .(7)$$

The coefficient $n_{0,f}e^2\lambda_r^2/m_0^*c^2$ has been absorbed into $\tilde{\Lambda}$. In addition, pulses that are only a few femtoseconds in duration that are initially broadband, may undergo strong phase modulation, compression, and frequency shifts. Combined with the fact that each successive power on the right-hand side of Eq.(7) generates progressively narrower functions in time with different frequency content and peak locations, a single-valued coefficient $\tilde{\Lambda}$ produced by expanding the denominator in Eq.(6) may not suffice to cover the vast wavelength range we consider. Therefore, we impart additional flexibility to the model by dropping the powers in $\tilde{\Lambda}$, and by adopting independent coefficients to write the final form of the free electron nonlinearity as follows:

$$\mathbf{P}_{f,NL} = -\tilde{\Lambda}_1(\mathbf{E}\cdot\mathbf{E})\mathbf{E} + \tilde{\Lambda}_2(\mathbf{E}\cdot\mathbf{E})^2\mathbf{E} - \tilde{\Lambda}_3(\mathbf{E}\cdot\mathbf{E})^3\mathbf{E} + \tilde{\Lambda}_4(\mathbf{E}\cdot\mathbf{E})^4\mathbf{E} . \qquad (8)$$

We note that beginning with Eq.(5), which assumes a nonlinear conductivity $\mathbf{J} = \sigma(\mathbf{E})\mathbf{E}$, the expansion of the denominator leads to a result that is practically identical to Eq.(8). Then, for two incident fields at the pump and probe carrier wave vectors and frequencies, $k_1$, $\omega_1$ and $k_2$, $\omega_2$, respectively, that are TM-polarized, the electric field **E** has components in the transverse (y) and longitudinal (z) directions, while the magnetic field **H** is polarized along the x-direction. The fields can be written as:

$$\mathbf{E} = \mathbf{j}\left(E_y^{\omega_1}(\mathbf{r},t)e^{i(k_1z-\omega_1 t)} + E_y^{\omega_2}(\mathbf{r},t)e^{i(k_2z-\omega_2 t)} + c.c.\right) + \mathbf{k}\left(E_z^{\omega_1}(\mathbf{r},t)e^{i(k_1z-\omega_1 t)} + E_z^{\omega_2}(\mathbf{r},t)e^{i(k_2z-\omega_2 t)} + c.c.\right) \quad (9)$$

$$\mathbf{H} = \mathbf{i}\left(H_x^{\omega_1}(\mathbf{r},t)e^{i(k_1z-\omega_1 t)} + H_x^{\omega_2}(\mathbf{r},t)e^{i(k_2z-\omega_2 t)} + c.c.\right) \quad . \qquad (10)$$

The bound electron polarization is expanded similarly to the electric field in Eq.(9). The carrier wave vectors and frequencies reflect pulses initially located in free space, propagating toward the sample. The field envelopes in equations (9) and (10) remain general functions of position and time. By maintaining all spatial and temporal derivatives in Maxwell's equations, we account for changes in the instantaneous phases and amplitudes of the fields and account for all orders of dispersion. Using the solutions of equations (2) and (3), the total polarization is the vector sum of the free and bound electron contributions, $\boldsymbol{P} = \boldsymbol{P}_f + \boldsymbol{P}_b$ which is inserted into Maxwell's equations. The effective mass $m_f^*(T_e)$ appears in both terms on the right-hand side of Eq.(2). However, only the respective linear terms survive when the expansion in Eq.(8) is applied to the nonlocal terms. To summarize the model described above, it currently accounts for hot electron generation; free carrier density increases, by shifting the nonlinearity from the denominator of the plasma frequency to its numerator [38]; nonlinear absorption; self- and cross-phase-modulation; and effects that occur in the sub-100 fs regime. Given the extremely short pulse durations, we presently do not account fully for carrier density relaxation and heat dissipation, although the equations of motion partially account for carrier dynamics and absorption. These issues will be tackled in future work.

Finally, we comment on the parameter space at our disposal, which is condensed into the expansions in Eqs. (4) and (8). One way to extract these parameters is to investigate harmonic generation [37, 38, 50]. For example, in [37], most of the generated harmonics are tuned in the free electron range. At low intensities, observing the conversion efficiencies of each harmonic in transmission and reflection leads to accurate determination of each of the coefficients in Eq.(8). In the case of silicon, considered in [38], at low intensities the observation of conversion efficiencies reveals the accuracy of the nonlinear Lorentzian coefficients defined above. At high intensity, the formation of plasma and free charge density activates the coefficients in Eq.(8). By accounting for conversion efficiencies at high and low intensities, it is possible to obtain a reasonably accurate representation of all the coefficients. However, in [3] the authors report only transmission and reflection of the probe under the action of the pump. Therefore, in the following we show that a set of parameters can be introduced to obtain good agreement with the experimental results in [3], enabling solid modeling of the physics at stake and predictions on what is possible when all aspects of the pump-probe nonlinear dynamics are brought to the fore.

**Results and Discussion: Transverse Plane Waves**

We begin our analysis by studying the experimental settings reported in [3] by assuming incident pulses having infinite transverse spatial extent, or transverse plane waves. As sketched in Fig.1, a 6 fs (2.2 optical cycles) pump pulse having carrier wavelength of 800 nm is normally incident on a 310 nm-thick ITO layer free-standing in vacuum. A probe pulse approximately 40 fs in duration and at least six orders of magnitude less intense than the pump pulse is tuned just below the ENZ crossing at 1200 nm and is incident at 2°. As detailed in the Supplementary Information, it is crucial to recognize that these pumping conditions are extreme. The spectrum of the pump pulse is sufficiently broad to overlap with the probe spectrum. Furthermore, the pump power density at the probe wavelength is comparable to or may even exceed the probe's power density, a condition commonly encountered in optical parametric amplifiers and, more broadly, in seeded nonlinear processes.

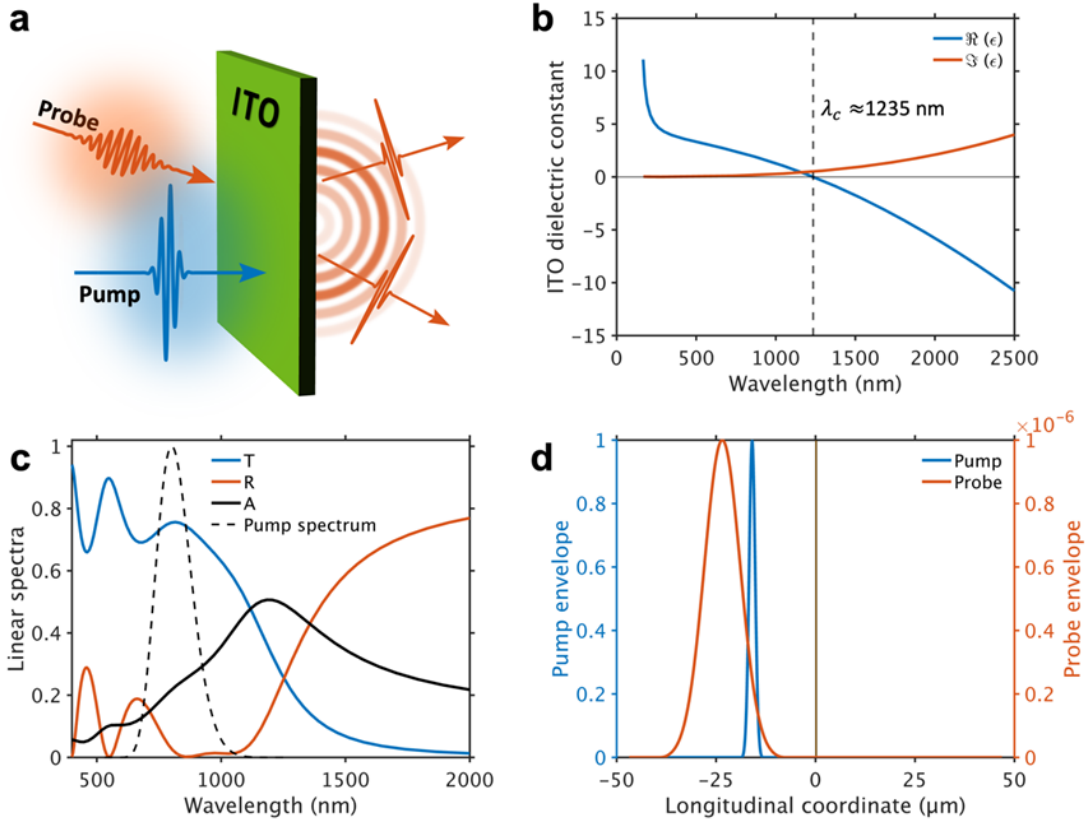

**Fig.1:** (a) Schematic representation of a strong pump and a weak probe incident on the ITO layer. (b) Local material dispersion of ITO having crossing point near 1235nm, represented by the vertical line. A Lorentzian response takes hold between 400nm and 500nm. The dispersion shown in Fig.1b is reproduced by inserting the following parameters in Eq.1: $\omega_{p,f} = 1.5785$, $\omega_{p,b} = 11.5$, $\gamma_f = 0.1105$, $\gamma_b = 0.01$, $\omega_{0,b} = 7$, with $\omega = 1/\lambda$, where $\lambda$ is in microns and all other parameters are scaled as indicated after Eq.(4). (c) Linear transmittance (T), reflectance (R), and absorption (A) from a free-standing, 310 nm-thick ITO layer surrounded by vacuum that displays the dispersion in Fig.1b for plane wave illumination. The spectrum of the 6 fs pulse is also shown (dashed black line). The etalon exhibits Fabry-Perot resonances below 800 nm that overlap with portions of the pump spectrum. (d) Spatial profiles of the initial locations of pump and probe envelope functions relative to each other's peak and to the layer's position. The probe peak power density (right axis) is at least six orders of magnitude smaller than the pump's peak power density (left axis).

In Fig. 2(a) we show the predicted probe transmittance and reflectance as functions of pump delay for peak power densities of 762 GW/cm², calculated based on our model described in the previous section. Mindful that this model accounts only for the fastest dynamical components of the free current density, the results of Fig. 2(a) nevertheless agree relatively well with the measurements reported in [3]. The reported measured reflectance in [3] displays a minimum at the same location as the transmittance, but never returns to the linear value, even after several hundred femtoseconds. If temperature, slower nonlinearities, enhanced absorption or interband transitions are affecting reflectance, then there is a reasonable expectation that they should also impact transmittance, which instead rapidly returns to its linear value, as in Fig. 2(a). Therefore, there are aspects of the dynamics that need to be further explored, to ascertain the reasons behind the behavior of the measured reflectance.

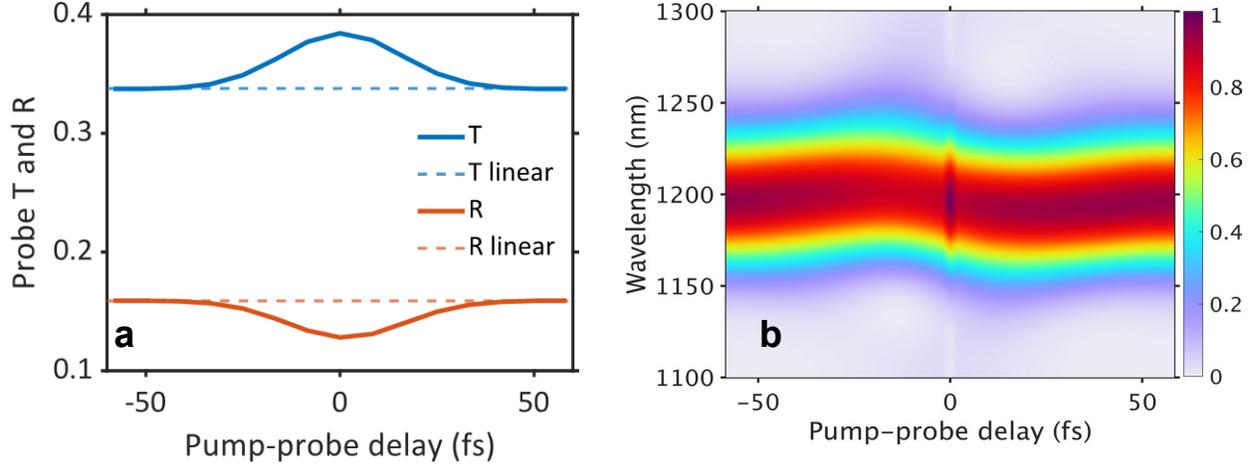

**Fig.2:** (a) Probe transmittance (T) and reflectance (R) as functions of pump delay. The peak power density of the pump is 762 GW/cm². $\beta = 3 \times 10^{-9}$, $\delta = 9 \times 10^{-18}$, $\vartheta = 2.7 \times 10^{-26}$, $\psi = 8.1 \times 10^{-35}$, $\tilde{\Lambda}_{\lambda,1} = 3 \times 10^{-10}$, $\tilde{\Lambda}_{\lambda,2} = 9 \times 10^{-20}$ $\tilde{\Lambda}_{\lambda,3} = 2.7 \times 10^{-29}$ $\tilde{\Lambda}_{\lambda,4} = 1.215 \times 10^{-39}$. (b) Probe spectrum vs pump delay. Negative delays redshift the spectrum, while positive delays lead to a blueshift. The results are consistent with experimental observations.

In Fig. 2(b) we plot the transmitted probe spectrum as a function of probe delay. The qualitative aspects once again agree well with experimental results in [3], with redshifts imparted by negative delays, and blueshifts resulting from positive delays. We note that the choice of a moderate peak power density and of the modeling coefficients is pivotal in the determination of the peak power density that is required to generate Fig. 2(a). In the Supplementary Information we show a version of Fig. 2 using a peak power density of 1.71 TW/cm², which triggers a regime where the probe pulse shows moderate gain and is compressed down to approximately 1 fs, resulting in supercontinuum generation.

Given the extremely short pulse durations we are considering, and the relative proximity of both pulses [see Fig.1(d)] all radiative processes cease before heat can be dissipated to any appreciable degree [49]. In contrast, currents can persist inside the layer after the pulses are gone, generating heat. We return to this issue below and in the Supplementary Information. The rate of decay is generally modified by a temperature-dependent damping term in Eq.(2), which should produce currents and number densities that decay accordingly. These effects will be considered in future work. Tuning conditions are such that the incident pump pulse spectrum sits at the edge of the Lorentzian range [Fig.1(c)]. Then, the results shown in Fig.2(a) are relatively insensitive to the magnitudes of the nonlinear coefficients in Eq.(4). For example, decreasing each of the Lorentzian coefficients by two orders of magnitude changes probe conversion efficiencies by approximately 1%. Therefore, free electron nonlinearities dominate the nonlinear dynamics, while the Lorentzian response is necessary to accurately reflect material dispersion across the spectrum.

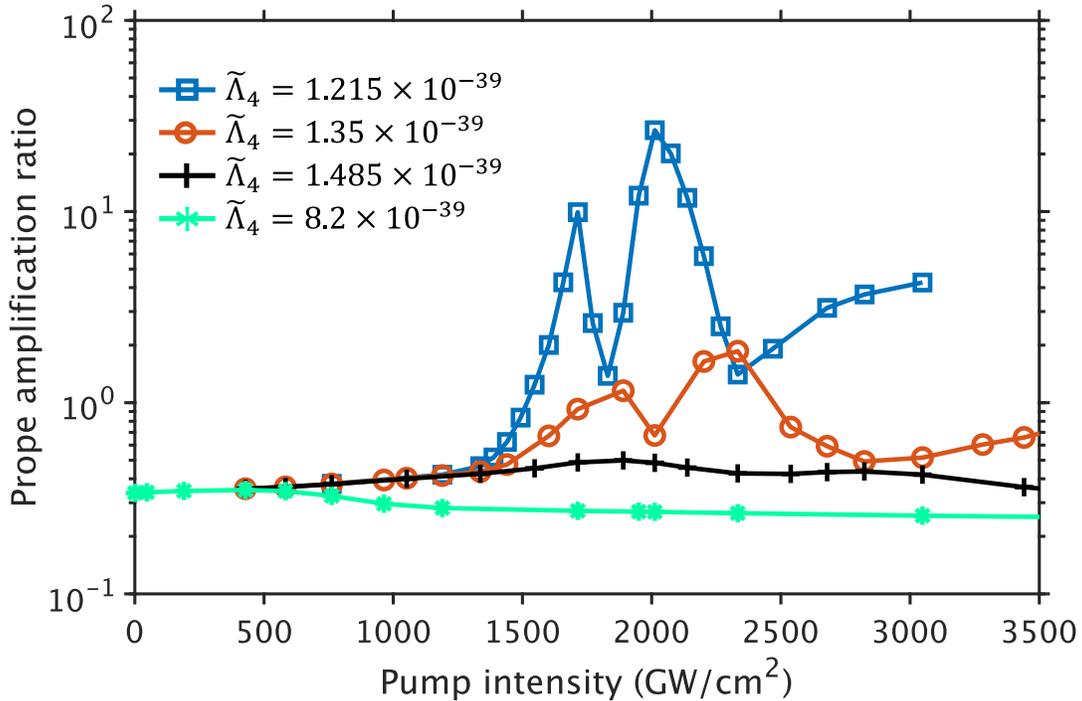

**Fig.3:** Predicted probe gain amplification factor in transmission as a function of pump peak power density and the magnitude of $\tilde{\Lambda}_4$. Probe reflection behaves in similar fashion. This quantity is calculated as the ratio of transmitted probe energy divided by initial probe energy.

As an instance pertaining to parameter sensitivity, we point to Fig.3 where we plot the *total transmitted probe energy* as a function of pump peak power density and the magnitude of $\tilde{\Lambda}_4$. The results suggest that for our choice of parameters, the probe transmitted energy is enhanced by nearly two orders of magnitude in resonant-like behavior near 2 TW/cm$^2$. This trend persists for thinner layers (see Supplementary Information for results on different layer thicknesses) and is thus somewhat insensitive to geometrical variations. The sensitivity of the results to the 9th order coefficient demonstrates that we are operating in an extreme pumping regime, where peak power densities suffice to activate all terms in the expansion. As noted earlier, given our general lack of knowledge about the actual magnitudes of the coefficients in Eq.(8) corresponding to the experimental values in [3], it is possible that a different set of parameters may also lead to similar results. The precise nature and magnitude of each coefficient should be assessed independently, for an accurate comparison with experimental results.

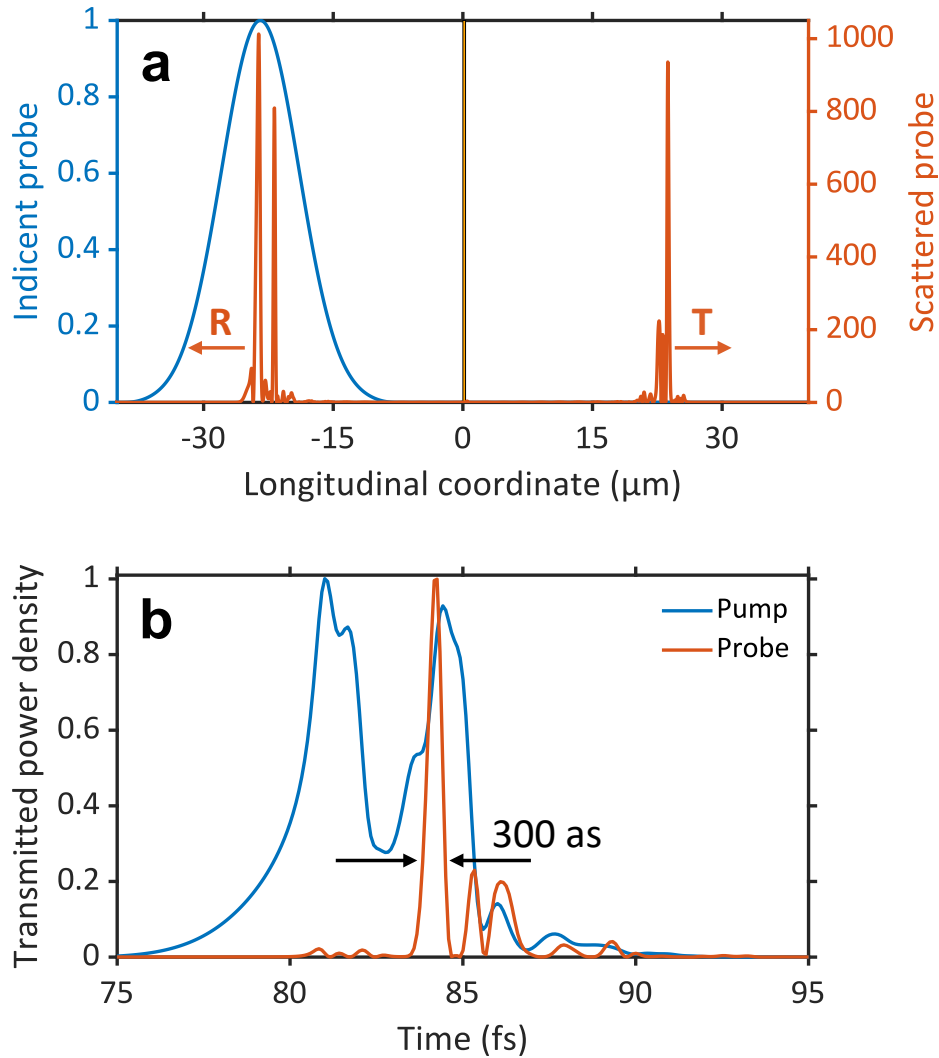

**Fig.4:** (a) Reflected and transmitted probe pulses generated following the interaction with a synchronized 6 fs, 2.072 TW/cm$^2$ pump pulse. All parameters are as in Fig.2(a). The arrows indicate direction of propagation. The incident probe is normalized and its amplitude plotted on the left axis. The peak power density of the probe is amplified by nearly three orders of magnitude. (b) Normalized temporal profiles of transmitted pump and probe pulses, as indicated in the figure. The probe forms a series of pulses whose FWHM is approximately 300 attoseconds, indicated by the arrows. The pump is also strongly modulated by the enormous nonlinear response.

A dramatic example of the extreme electrodynamics that unfolds in this regime may be learned by visualizing spatial and temporal field profiles in the high gain region. In Fig.4(a) we show the spatial profiles of incident and scattered fields for synchronized incident pump-probe pulses, when the peak power density of the incident pump is 2.072 TW/cm$^2$. We also show the normalized incident probe field intensity. The amplitudes of the scattered fields are normalized with respect to the peak power density of the incident probe pulse. In addition to the predicted extreme spatial compression, the peak power density of reflected and transmitted probe fields are amplified by nearly three orders of magnitude. In Fig.4(b) we collect and plot the field intensities as the transmitted pulses sweep past an observation point located to the right of the layer. The spatial compression of the probe pulses illustrated in Fig.4(a) corresponds to probe pulses having a full width at half maximum (FWHM) of 300 attoseconds, or 1/10 of an optical cycle at 1.2 μm, shown in Fig.4(b). Similar pulse durations are generated in reflection.

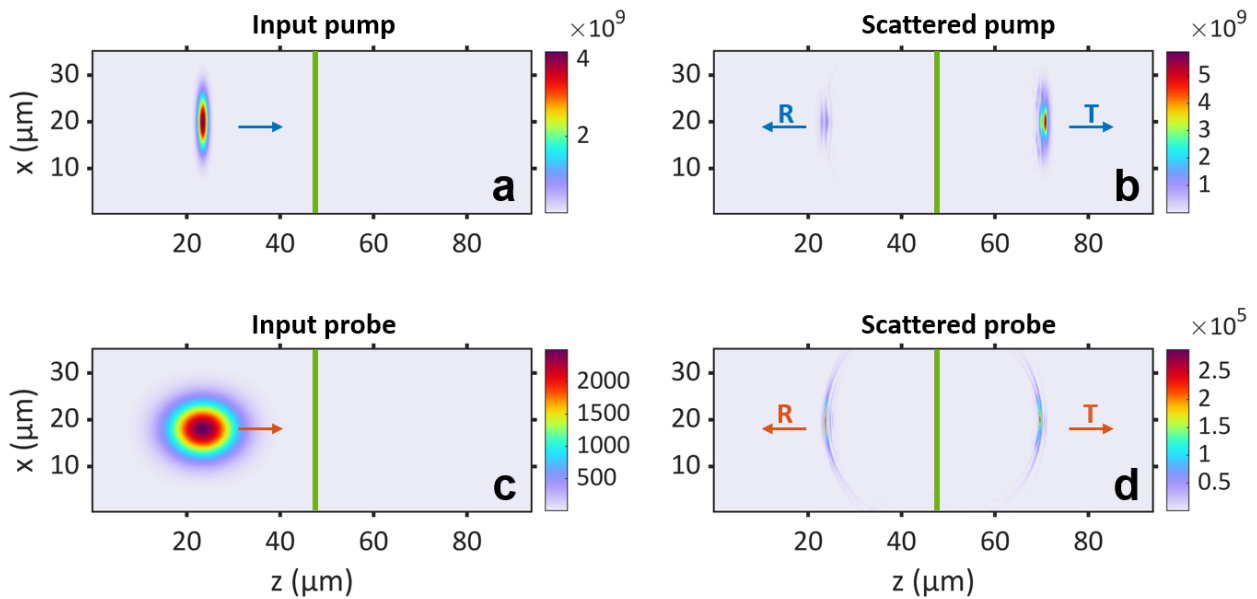

**Fig.5:** (a) Incident pump pulse. Initial longitudinal spatial extension of the pump is approximately 1.8 μm. (b) Transmitted and reflected pump pulses. Both scattered pulses form multiple peaks, not unlike those shown in Fig.4(b). A close look at the intensity profiles reveals curvature due to diffraction. (c) Incident, 40 fs probe pulse. Transverse beam widths are approximately 10 μm for both beams, resulting in little or no diffraction of either pulse if the propagation proceeded in free space for the specified distances. (d) Scattered probe pulses form nearly perfect spherical waves as they recede from the ITO layer. The attached video1 details the interaction from beginning to end.

**Strong Diffractive Coupling Regime**

In this section, we investigate the effects of diffraction by examining pulses with finite transverse widths. When considering linear or free-space propagation, a thickness of 310 nm represents only a small fraction of the wavelength, meaning that significant diffraction effects are not expected for most beam widths. However, if the peak power densities approach and exceed 1 TW/cm². both field curvature and diffractive coupling can intensify dramatically. In Fig.5, we present incident and scattered pump and probe field intensities, based on the same conditions and parameters used in Fig.2(a). The results in Fig.5 clearly demonstrate that the interaction between the pulses is substantial for both the pump and probe, showing significant diffractive effects. The pump pulse begins to break up, and its intensity profile shows signs of curvature, a result of diffraction effects. In contrast, when the probe exits the medium in either direction, it diffracts to such an extent that it appears as if it originated from a point source. The attached video1 details the entire interaction.

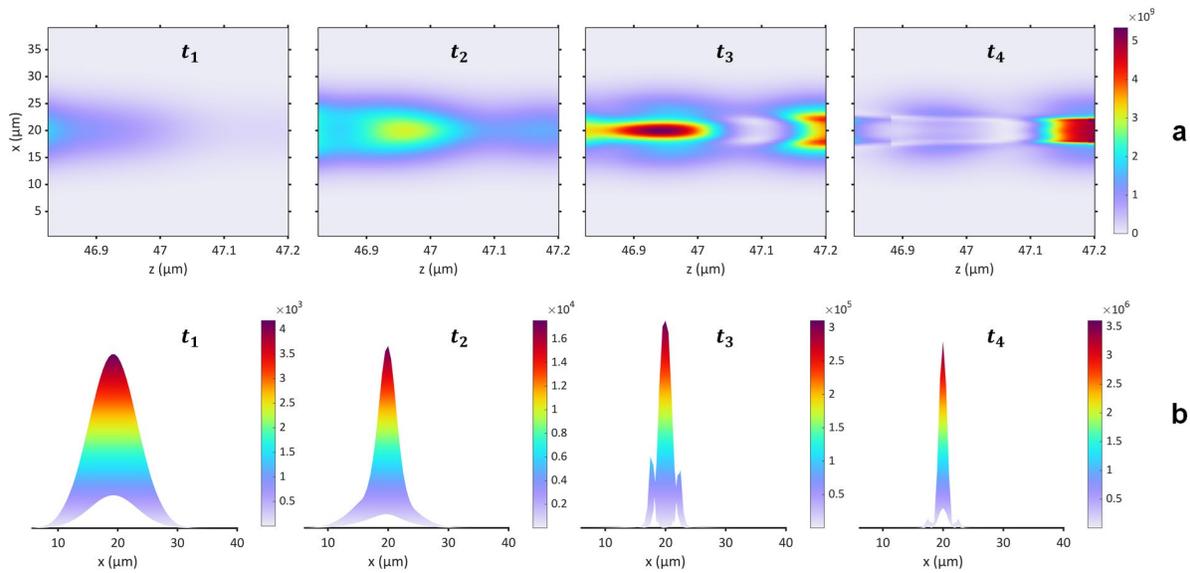

**Fig.6:** Top Row (a) Consecutive pump snapshots as it enters the ITO layer from the left. Self-focusing and self-guiding is evident. Bottom Row (b) Same as (a) but for the probe field intensity, as seen from the exit side. The probe's *entire transverse field profiles* narrow considerably from 10μm down to less than 1μm, following the extraordinary degree of focusing experienced by interacting with the pump. Subsequent snapshots show that both fields resonate inside the ITO layer.

Our simulations show that if the pump pulse is a transverse plane wave, this phenomenon does not occur, regardless of probe shape. The pronounced diffraction of the probe that we predict can be triggered *only if the pump pulse has a finite transverse profile and is allowed to undergo self-focusing within the medium*. Focusing and longitudinal waveguiding of the pump pulse inside the layer are evident in the top row (a) of Fig.6, where we show four consecutive snapshots that track the pump's ingress into the medium. In this sequence, the pump is seen to focus and form a conduit that not only guides but also compresses the probe into a channel only a few microns in width. Upon egress, the pump diffracts. The bottom row (b) of Fig.6 tracks the front view of the probe's transverse profile as it moves along the channel at each of the pump's snapshot times, as indicated in the figure. By snapshot t₄, bottom row, *the probe's entire transverse profile has been compressed to just 1 μm (or 0.82 λ), along the entire channel*. This strong, subwavelength confinement of the probe is eventually overcome, leading to the dramatic diffraction pattern observed in Fig.6, as the probe field exits the medium.

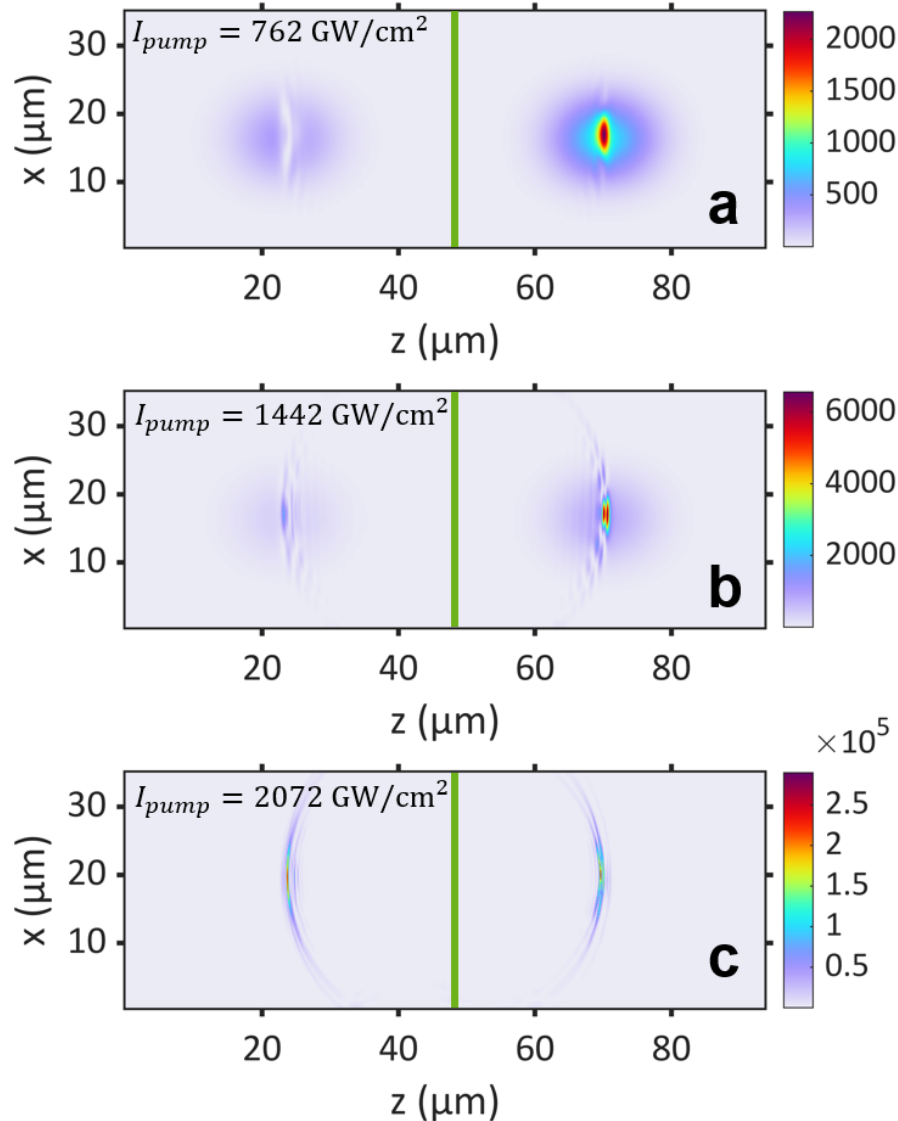

**Fig.7:** Transmitted and reflected probe pulses for peak power density of 762 GW/cm² (a); 1440 GW/cm² (b); 2072 GW/cm² (c), identical to Fig.5d.

Finally, in Fig.7 we plot the scattered probe fields for two intermediate pump peak power densities and reproduce Fig.5(d) for direct comparison. The figures suggest that the onset of the waveguiding mechanism at high intensities is clearly the trigger of the spherical wave that disperses probe energy across the transverse direction.

**Time-Varying Electromagnetic Response of the Material**

As discussed and demonstrated above, neglecting pump dynamics in favor of an arbitrary, time-dependent electromagnetic response greatly oversimplifies the problem, possibly concealing the most important, hitherto unknown aspects. The assumption that an equation of the form $D(z,t) = \varepsilon(z,t)E(z,t)$ can be used to model the pump, absent a rigorous formal derivation, is generally flawed. However, given the general tendency to use this approach to model the problem, we now attempt to reconcile our results with the current literature, with the understanding that some significant assumptions are necessary to introduce a time-varying complex permittivity. We begin by expanding the free and bound polarization densities, as well as the electric field as:

$$\mathbf{A}(\mathbf{r},t) = \mathbf{A}_1(\mathbf{r},t)e^{-i(\omega_1 t - k_1 z)} + \mathbf{A}_2(\mathbf{r},t)e^{-i(\omega_2 t - k_2 z)} + c.c. \quad (11)$$

where the vector $\mathbf{A}_1(\mathbf{r},t)$ is the complex envelope of the pump field, oscillating at the carrier frequency $\omega_1$ and spatial frequency $k_1$, $\mathbf{A}_2(\mathbf{r},t)$ is the complex envelope of the probe field, oscillating at the carrier frequency $\omega_2$ and spatial frequency $k_2$. The quantity $\mathbf{A}$ may denote electric field $\mathbf{E}$, the free-electron polarization density $\mathbf{P}_f$, or the bound-electron polarization density $\mathbf{P}_b$. We assume that the pump field dynamics remain largely unaffected by the probe field, an approximation justified by the cases under investigation, where the pump intensity significantly exceeds that of the probe. *Neglecting nonlocal terms,* which generally cross-couple polarization components, and substituting Eq. (11) into Eqs. (2) and (3), yields two distinct equations of motion for the complex envelopes of the probe field: one governing free electrons,

$$\frac{\partial^2 \mathbf{P}_{f,2}(\mathbf{r},t)}{\partial t^2} + (\gamma_f - 2i\omega_2)\frac{\partial \mathbf{P}_{f,2}(\mathbf{r},t)}{\partial t} - (\omega_2^2 + i\omega_2\gamma_f)\mathbf{P}_{f,2}(\mathbf{r},t) = \epsilon_0 \omega_{p,f}^2(\mathbf{r},t)\mathbf{E}_2(\mathbf{r},t) \quad (12)$$

and another governing bound electrons,

$$\frac{\partial^2 \mathbf{P}_{b,2}(\mathbf{r},t)}{\partial t^2} + (\gamma_b - 2i\omega_2)\frac{\partial \mathbf{P}_{b,2}(\mathbf{r},t)}{\partial t} - [\omega_2^2 + i\omega_2\gamma_f - \omega_b^2(\mathbf{r},t)]\mathbf{P}_{b,2}(\mathbf{r},t) = \epsilon_0 \omega_{p,b}^2 \mathbf{E}_2(\mathbf{r},t). \quad (13)$$

In Eq. (12) and Eq. (13), we have introduced the space-time-dependent plasma frequency for free electrons:

$$\omega_{p,f}^2(\mathbf{r},t) = \omega_{p,f}^2 - \Lambda_1|\mathbf{E}_1(\mathbf{r},t)|^2 + \Lambda_2|\mathbf{E}_1(\mathbf{r},t)|^4 - \Lambda_3|\mathbf{E}_1(\mathbf{r},t)|^6 + \Lambda_4|\mathbf{E}_1(\mathbf{r},t)|^8 \quad (14)$$

Similarly, we define the space-time-dependent resonance frequency of bound electrons:

$$\omega_{0,b}^2(\mathbf{r},t) = \omega_{0,b}^2 - \beta_1|\mathbf{P}_{b,1}(\mathbf{r},t)|^2 + \beta_2|\mathbf{P}_{b,1}(\mathbf{r},t)|^4 - \beta_3|\mathbf{P}_{b,1}(\mathbf{r},t)|^6 + \beta_4|\mathbf{P}_{b,1}(\mathbf{r},t)|^8. \quad (15)$$

*The frequency expressions in Eqs.(14-15) are derived by considering only the nonlinear cross-phase modulation terms proportional to the product of the pump field intensity $|E_1|^2$ and the vector*

*field $E_2$, and by neglecting source-type nonlinear polarization terms associated with frequency mixing.* In video2 and video3 we show the temporal evolutions of $\delta\omega_{p,f}^2(\mathbf{r},t) = \frac{\omega_{p,f}^2(\mathbf{r},t) - \omega_{p,f}^2}{\omega_{p,f}^2}$ and $\delta\omega_{0,b}^2(\mathbf{r},t) = \frac{\omega_{0,b}^2(\mathbf{r},t) - \omega_0^2}{\omega_0^2}$, respectively, as a 1.7 TW/cm², 6 fs pump pulse sweeps across the ITO layer. While the plasma frequency is modulated by as much as ~50% of its initial value, the resonance frequency alteration is of order 1%, thus demonstrating the impact of the pump and the importance of properly accounting for material dispersion across the spectrum. Treating the space-time-dependent plasma frequency $\omega_{p,f}^2(\mathbf{r},t)$ and the space-time-dependent resonance frequency $\omega_b^2(\mathbf{r},t)$, as mere parameters in complex Drude and Lorentz permittivity models, respectively, is neither as straightforward nor as accurate as it might seem. In the quasi-monochromatic approximation for the probe field, the time-derivatives of the complex envelopes appearing in Eq.(12) and Eq.(13) can be neglected. Therefore, only then may one introduce a space-time-dependent complex permittivity:

$$\epsilon(\mathbf{r},t) = 1 - \frac{\omega_{p,f}^2(\mathbf{r},t)}{\omega^2 + i\omega\gamma_f} - \frac{\omega_{p,b}^2}{\omega^2 + i\omega\gamma_b - \omega_b^2(\mathbf{r},t)} \qquad (16)$$

This function relates the quasi-monochromatic probe field $\tilde{\mathbf{E}}_2(\mathbf{r},t)$ to its polarization field $\tilde{\mathbf{P}}_{f,2}(\mathbf{r},t) + \tilde{\mathbf{P}}_{b,2}(\mathbf{r},t)$, i.e. $\tilde{\mathbf{D}}_2(\mathbf{r},t) = \tilde{\mathbf{E}}_2(\mathbf{r},t) + 4\pi\left(\tilde{\mathbf{P}}_{f,2}(\mathbf{r},t) + \tilde{\mathbf{P}}_{b,2}(\mathbf{r},t)\right) = \varepsilon(\mathbf{r},t)\tilde{\mathbf{E}}_2(\mathbf{r},t)$.

While the complex permittivity in Eq. (16), incorporating the Drude-Lorentz parameters from Eqs. (14) and (15), accurately captures the strong, high-order self-phase modulation of the pump – a feature often overlooked in the literature – this model may yet fail to account for critical aspects of the problem in several ways: (i) At high intensities pump dynamics should be included; (ii) The monochromatic approximation may not always be valid, even when the probe input field is harmonic. This is due to the rapid temporal variation of the free-electron plasma and bound-electron resonance frequencies, which, in the scenario under investigation, occurs on a timescale of just a few femtoseconds; (iii) The pump spectrum overlaps with the probe spectrum, potentially leading to undesired spectral mixing effects; (iv) The probe field is broadband, challenging the assumption of a single-frequency response used to arrive at Eq.(16); (v) Additional nonlinear terms emerge in Eqs. (14) and (15) when the probe intensity becomes comparable to the pump intensity, further complicating the system's dynamics; (vi) Nonlocal effects can be determining factors in thinner layers, and their neglect can also lead to misleading results. All these factors should be considered when attempting to model extreme electrodynamics conditions using simplistic approaches.

**Summary and Conclusions**

We have performed a theoretical analysis of a pump-probe interaction upon a TCO thin film which displays extreme electrodynamic conditions. We apply our model to the experimental results reported in [3], where high intensities and two-cycle pulses pump an ITO layer. Our results suggest that maximum transmittance and minimum reflectance occur when the pump and probe pulses are synchronized, with the results being sensitive to free electron nonlinearities. Additionally, for the chosen parameters, which are neither optimized nor exhaustive in any way, a threshold is identified where resonant probe gain near 2 TW/cm² is favored, for thick and thin layers alike, suggesting that this behavior is relatively independent of geometrical factors. In the high-gain region, both reflected and transmitted probe peak power densities are amplified by nearly three orders of magnitude. The spatial compression of the probe pulse results in durations as short as 300 attoseconds, or 1/10 of one optical cycle at the probe's wavelength, with similar compression occurring in reflected pulses, highlighting a novel pathway to the attosecond regime and supercontinuum generation in deeply subwavelength material thicknesses. In what amounts to an attempt to reconcile our results with approaches prevalent in the literature, we have derived a complex, spatio-temporal permittivity under the conditions and constraints enumerated in the last section. Our analysis demonstrates that the spatio-temporal dynamics of the dielectric response cannot be reduced to any simple analytic form and is neither a real function nor sinusoidal during the interaction. We have also examined the effects of diffraction on pulses having finite transverse widths, noting that diffraction becomes significant even for subwavelength propagation lengths when peak power densities exceed 1 TW/cm². At high intensities, the pump pulse self-focuses within the medium, creating a narrow waveguide that compresses the probe pulse along the transverse direction down to approximately 1 µm in width. This narrow, subwavelength beam eventually overcomes confinement and exhibits diffraction as if it originated from a point source. Pulses of longer duration will experience the effects of temperature, which in turn can affect damping rates as well as nonlocal effects. Finally, we find relatively good qualitative agreement between the experimental results reported in reference [3], and the results we have reported above. These results highlight the need for advanced material modelling when an optical excitation occurs under extreme conditions in terms of pulse bandwidth and peak intensities, circumstances that are currently trending in time-varying systems.

## Acknowledgments

M.F. wishes to acknowledge economic support from EPSRC project ID: EP/X035158/1, AFOSR (EOARD) under Award No. FA8655-23-1-7254. C.R. acknowledges partial support by the European Union - NextGenerationEU, Mission 4, Component 1, under the Italian Ministry of University and Research (MUR) National Innovation Ecosystem grant ECS00000041 - VITALITY - CUP E13C22001060006. A.A. acknowledges support from the Department of Defense and the Simons Foundation. The authors wish to thank M. Clerici, G. Della Valle, A. Marini, C. De Angelis, and Y. Sivan for critical reading of the manuscript.